%% file: arxiv_main_paper_final.tex
\documentclass[%
 reprint,
superscriptaddress,
nobibnotes,
 amsmath,amssymb,
prl,
floatfix,
]{revtex4-2}

\usepackage{graphicx}
\usepackage{dcolumn}
\usepackage{bm}
\usepackage{siunitx}
\usepackage{comment}
\usepackage{import}
\usepackage{MnSymbol}
\usepackage{xcolor}
\usepackage[normalem]{ulem}
\usepackage[sort&compress]{natbib}
\usepackage{bibunits}
\usepackage[caption=false]{subfig}

\usepackage[linktocpage=true]{hyperref}
\hypersetup{  colorlinks, 
              linktoc         = page, 
              linkcolor       = {orange!60!black}, 
              urlcolor        = {cyan!60!black}, 
              citecolor       = {cyan!60!black}, 
              anchorcolor     = {yellow}
}

\makeatletter
\renewcommand{\fnum@figure}{\textbf{Fig.~\thefigure}}
\makeatother
\usepackage{etoolbox}

\def\urlprefix{}

\begin{document}

\preprint{Draft 1}

\title{Primary thermalisation mechanism of Early Universe observed from Faraday-wave scattering on liquid-liquid interfaces}

\author{Vitor S. Barroso}%
\email[E-mail: ]{vitor.barrososilveira@nottingham.ac.uk}
\affiliation{School of Mathematical Sciences, University of Nottingham, University Park, Nottingham, NG7 2RD, UK}
\affiliation{Centre for the Mathematics and Theoretical Physics of Quantum Non-Equilibrium Systems, University of Nottingham, Nottingham, NG7 2RD, UK}
\affiliation{Nottingham Centre of Gravity, University of Nottingham, Nottingham, NG7 2RD, UK}

\author{August Geelmuyden}
\email[E-mail: ]{august.geelmuyden@nottingham.ac.uk}
\affiliation{School of Mathematical Sciences, University of Nottingham, University Park, Nottingham, NG7 2RD, UK}
\affiliation{Centre for the Mathematics and Theoretical Physics of Quantum Non-Equilibrium Systems, University of Nottingham, Nottingham, NG7 2RD, UK}
\affiliation{Nottingham Centre of Gravity, University of Nottingham, Nottingham, NG7 2RD, UK}

\author{Zack Fifer}
\affiliation{School of Mathematical Sciences, University of Nottingham, University Park, Nottingham, NG7 2RD, UK}
\affiliation{Centre for the Mathematics and Theoretical Physics of Quantum Non-Equilibrium Systems, University of Nottingham, Nottingham, NG7 2RD, UK}
\affiliation{Nottingham Centre of Gravity, University of Nottingham, Nottingham, NG7 2RD, UK}

\author{Sebastian Erne}
\affiliation{School of Mathematical Sciences, University of Nottingham, University Park, Nottingham, NG7 2RD, UK}
\affiliation{Vienna Center for Quantum Science and Technology, Atominstitut, TU Wien, Stadionallee 2, 1020 Vienna, Austria}

\author{Anastasios Avgoustidis}
\affiliation{
School of Physics and Astronomy, University of Nottingham, Nottingham NG7 2RD, United Kingdom
}
\affiliation{Nottingham Centre of Gravity, University of Nottingham, Nottingham, NG7 2RD, UK}

\author{Richard J. A. Hill}
\affiliation{
School of Physics and Astronomy, University of Nottingham, Nottingham NG7 2RD, United Kingdom
}
\author{Silke Weinfurtner}
\affiliation{School of Mathematical Sciences, University of Nottingham, University Park, Nottingham, NG7 2RD, UK}
\affiliation{Centre for the Mathematics and Theoretical Physics of Quantum Non-Equilibrium Systems, University of Nottingham, Nottingham, NG7 2RD, UK}
\affiliation{Nottingham Centre of Gravity, University of Nottingham, Nottingham, NG7 2RD, UK}

\begin{abstract}
    For the past two hundred years, parametric instabilities have been studied in various physical systems, such as fluids, mechanical devices and even inflationary cosmology. It was not until a few decades ago that this subharmonic unstable response arose as a central mechanism for the thermalisation of the Early Universe, in a theory known as preheating. Here we study a parametrically driven two-fluid interface to simulate the key aspects of inflationary preheating dynamics through the onset of nonlinear Faraday waves. We present a detailed analysis of the effective field theory description for interfacial waves through the factorization properties of higher-order correlations. Despite the intricacies of a damped and highly interacting hydrodynamical system, we show that the scattering of large amplitude Faraday waves is connected to a broadening of primary resonance bands and the subsequent appearance of secondary instabilities as predicted in preheating dynamics.
\end{abstract}

\maketitle
\defaultbibliography{arxiv_main_paper_final}
\defaultbibliographystyle{naturemag}
\begin{bibunit}

Parametric instabilities can be responsible for dramatic events, from the collapse of bridges~\cite{Gazzola2015BriefBridges} and rolling of ships at sea~\cite{Biran2014ChapterWaves} to the thermalisation of our universe following cosmic inflation, 13.8 billion years ago~\cite{Kofman:1994rk,Shtanov:1994ce,Kofman:1997yn}. In a leading theory for the thermalisation of the Early Universe,
known as preheating, broad parametric resonance efficiently transfers the energy of the inflaton field to other fields and particles, thus producing the hot plasma required for the Big Bang theory to proceed. However, direct observations of the non-linear dynamics of preheating in the early universe are not feasible. Here, we conduct a controlled experiment to simulate the key aspects of inflationary preheating in a parametrically driven interface between two fluids~\cite{Faraday1831XVII.Surfaces,Miles90}. We study the scattering of large amplitude Faraday waves and observe a broadening of primary resonance bands and the subsequent appearance of secondary instabilities and their estimated growth rates~\cite{Berges2003ParametricTheory,Zache2017InflationaryCondensates}, as predicted in preheating. Adapting the statistical machinery from field theories, namely two-point functions and the factorisation properties of higher order correlators~\cite{Schweigler2017Experimental,Zache20_1PI,Prufer2020Experimental}, we show that the interfacial evolution is accurately captured by leading terms in an effective perturbative description. Our results demonstrate the robustness of preheating dynamics in a strongly interacting and damped system.

The study of instabilities on fluid surfaces caused by an external vibration dates back to $1831$, when Faraday first detailed the phenomenon in a cylindrical glass filled with water~\cite{Faraday1831XVII.Surfaces}. He noted that the unstable waves arising on the fluid surface oscillate with half of the external driving frequency. This sub-harmonic response characterises parametric resonance, which, since then, has been extensively examined and identified in a wide range of physical systems~\cite{Kovacic2012SpecialEngineering:}.  In general, when subjected to a periodic forcing with a frequency of $\omega_\mathrm{d}$, the spectral response of these systems displays the so-called unstable resonance bands at specific frequencies~\cite{Kovacic2018MathieusFeatures}. For a common form of the coupling that we consider here, the unstable bands occur at integer multiples of $\omega_0\equiv\omega_\mathrm{d}/2$. Within such resonance bands, parametric amplification happens with a common exponential rate but is dominated by the primary instability at $\omega_0$. As amplitudes grow larger, non-linear effects come into play and limit the amplification.

The interface between two fluids undergoing a vertical oscillatory acceleration is no different to Faraday's original system and displays a similar unstable behaviour~\cite{Kumar94}. While two coupled sets of Navier-Stokes equations govern the fluids' motion, interfacial waves evolve according to an effective set of dynamical equations. Fifer et al.~\cite{Fifer2019AnalogField} showed that this emergent description is able to emulate the propagation of scalar fields in various cosmological scenarios. This is one of many systems, where small excitations in a fluid or superfluid, such as sound or interface waves, experience an effective spacetime geometry~\cite{barcelo2011analogue} provided by the fluid flow.
From an experimental standpoint, gravity simulators have been very successful in mimicking a variety of black hole (e.g.~Hawking radiation~\cite{Weinfurtner2011MeasurementSystem,Euve16HawkingWater,munoz2019observation,Drori19OptHawking,kolobov2021observation}, superradiance~\cite{Torres2017RotationalFlow,Cromb2020Amplification,Braidotti22Penrose} and ringdown~\cite{Torres20QNM}) and cosmological (e.g.~Hubble friction~\cite{Eckel2018ExpandingBEC} and excitations in  Friedmann-Roberston-Walker-type spacetimes~\cite{Jaskula12DynamicalCasimir,Schuetzhold07IonTrap,Prain17,steinhauer2022analogue,viermannQuantumFieldSimulator2022a}) processes in the lab. As such, gravity simulators open the possibility of probing fundamental processes of otherwise inaccessible physical systems. These simulators rely on the experimental study of a physically realisable ``analogue system" that has, to some approximation, the same mathematical description as the fundamental phenomena of interest.

\begin{figure*}[ht!]
\centering
\makebox[0pt]{\includegraphics[width = \linewidth]{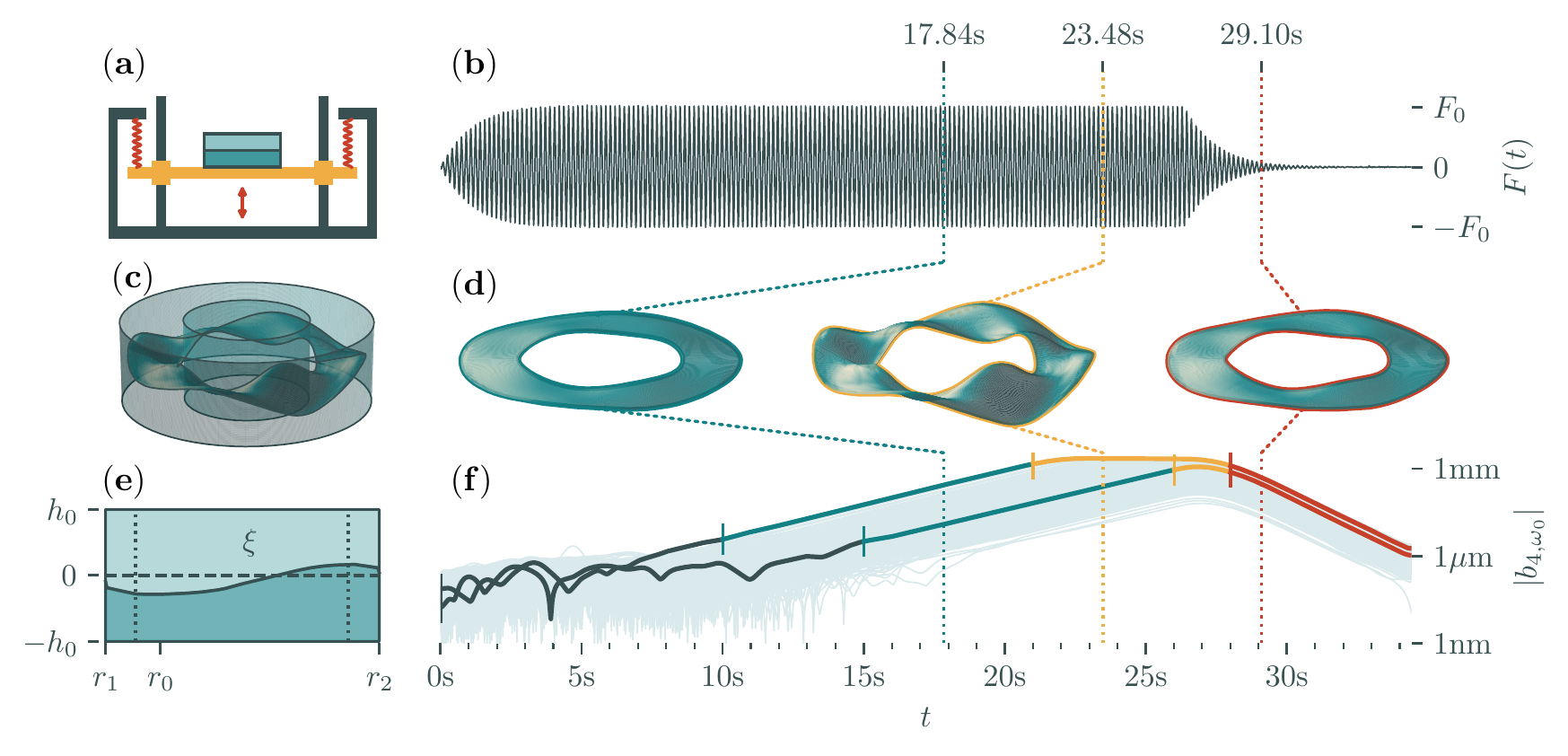}}
\caption{\textbf{Experimental setup and interfacial dynamics.}
\textbf{(a)}: Schematic depiction of the driving setup. The fluid cell (blue rectangles) sits on a platform (in orange) supported by springs (in red) and guided by vertical rods with pneumatic air-bearings. The up-down red arrow illustrates the harmonic motion of the system. \textbf{(b)}: A typical profile of the measured vertical acceleration $F(t)$ over time. The driver starts at $t=0$ and stops at $t=26.36$s. Damping in the spring-mass driving platform creates a transient amplitude change in $F(t)$, which takes about $4$ seconds to reach a constant value. \textbf{(c)}: Inset of \textbf{(a)} portraying the geometry of the fluid cell with the experimentally reconstructed interface at the instance ($t=23.48$s) of maximal amplitude.
\textbf{(d)}: Three rendered temporal snapshots of the experimentally observed interfacial waves. \textbf{(e)}: Cross-section of the fluid cell in \textbf{(c)} at fixed azimuthal angle $\theta_0$ portraying the interfacial height $z=\xi(t,r,\theta_0)$ in comparison with the horizontal walls at $\pm h_0 = \pm 17.5mm$. When at rest, the surface is located at $z=0$. The dashed, vertical lines indicate the radii where the steep curvature of the menisci between the interface and the vertical walls at $r_1=20\mathrm{mm}$ and $r_2=40\mathrm{mm}$ cause the detection method to fail. \textbf{(f)}: Instantaneous amplitudes $b_{m,\omega_0}$ in logarithmic scale of the azimuthal mode $m=4$, at the primary resonance frequency $\omega_0$, for $1500$ repetitions (light blue) at fixed radius $r_0=24\mathrm{mm}$. Out of the ensemble of repetitions, two qualitatively distinct runs are highlighted in bold with colored regions showing different parts of the dynamics: (dark green) detection method noise floor, (turquoise) log-linear unstable growth, (orange) transition from amplification-dominated to damping-dominated dynamics and (red) log-linear unstable decay. The difference in the two realisations is due to the randomness of the initial state.
}
\label{fig:setup_evol}
\end{figure*}

In this context, the powerful machinery of Effective Field Theories (EFT) becomes extremely useful, allowing one to classify, and systematically study, non-linear correction terms in the effective descriptions of both the fundamental system and its experimental analogue. Generally, the EFT descriptions of the two systems are not identical, in that higher-order corrections differ. However, one often observes that the expected phenomenology still arises in the experimental analogue, thus demonstrating the universality and robustness of the physical phenomena under study. Moreover, this experimental approach allows one to tune the system's configurations and run the experiment repetitively, in effect scanning the parameter space of the EFT. The bridge between the experiment and the fundamental system is the statistical machinery of EFT, particularly, correlation functions. The latter can be readily computed in the effective description, and also accurately reconstructed from direct observations of the experimental analogue system, as demonstrated in pioneering work on ultra-cold atoms systems~\cite{Schweigler2017Experimental,Zache20_1PI,Prufer2020Experimental,Bloch2012QuantumGases}.

Here, we investigate driven parametric instabilities on a liquid-liquid interface and observe their subsequent non-linear breakdown, which exhibits key features of preheating, the leading theory for the thermalisation of the early Universe. We employ methods introduced in~\cite{Schweigler2017Experimental,Zache20_1PI} to develop and validate the EFT of interfacial mode-mode interactions in our system. Hence, the statistical machinery can provide a tool to extend the programme of gravity simulators from free to interacting fields, allowing us to explore intrinsically non-linear scenarios~\cite{Bloch2012QuantumGases}, such as cosmic preheating~\cite{Zache2017InflationaryCondensates}, experimentally.

We perform a series of automated and synchronised parametric resonance experiments on the interface of a biphasic solution of potassium carbonate, ethanol and water enclosed in an annular cylindrical cell (see Supplementary Information for details). A driving platform (Fig.~\ref{fig:setup_evol} (a)) oscillates the fluid cell vertically with measured acceleration $F(t)$ (Fig.~\ref{fig:setup_evol} (b)), whose frequency and amplitude are set to $\omega_{\mathrm{d}}/(2\pi)=  6.07 ~ \mathrm{Hz}$ and $F_0=0.352 g=3.45 ~ \mathrm{m}~\mathrm{s}^{-2}$, respectively. During each repetition, we observe the unstable evolution of the interfacial elevation $\xi$ reconstructed from the experiment (Fig.~\ref{fig:setup_evol} (c), see Supplementary Information). As depicted in Fig.~\ref{fig:setup_evol} (d), the resonant modes appear as azimuthal waves on the interface, i.e., fixed $m$ values in the decomposition $\xi(t,r,\theta)=\sum_m\xi_m(t,r)\exp(im\theta)$. Their radial profile and the cell's radial cross-section are shown in Fig.~\ref{fig:setup_evol} (e).

We observe that, throughout the system's evolution, the azimuthal mode $m=4$ is the dominant parametrically amplified wave, as seen by the four crests on the interface in Fig.~\ref{fig:setup_evol} (d).
The essential features of this mode's evolution around the primary resonance band at $\omega_0$ are captured by the time-dependent complex envelope $b_{m=4,\omega_0}$, defined from the frequency decomposition of the azimuthal mode, $\xi_m\propto \int d\omega [b_{m,\omega}e^{-i \omega t} + b_{-m,\omega}^*e^{i \omega t}]$. These instantaneous amplitudes are displayed in logarithmic scale in Fig.~\ref{fig:setup_evol} (f) for all experimental repetitions (in light blue), and two qualitatively distinct realisations are highlighted to stress distinguishable stages of the evolution.

Initially, we see the detection scheme's noise level (in dark green) overwhelming the signal until the instabilities grow out of it between $10$ and $17$ seconds. A clear log-linear trend (turquoise region) appears in all repetitions with consistent exponential growth rates throughout. Similarly, all runs are seen to decay exponentially (in red) after the driver is turned off at $26.3$ seconds. However, while some repetitions transition quickly from the amplifying region to the final damping stage (orange region), others saturate in amplitude before that. With the latter case, see upper curve in Fig.~\ref{fig:setup_evol} (f), we observe a bound to the growth of the unstable interfacial waves, which is inconsistent with a single-mode linear evolution and indicates that the energy continuously provided by the driver is being scattered into other modes in the system~\cite{Edwards1994PatternsExperiment,Zhang1996SquareWaves,Zhang1997PatternWaves,Chen1999AmplitudeWaves,Chen2002NonlinearInstabilities,Garih2013OnInstability,Ciliberto1985ChaoticWaves}. In the early universe, the driver effectively shuts off due to backreaction from produced particles, which drains energy out of the inflaton condensate and dampens the oscillations that had driven parametric resonance~\cite{Amin:2014eta}.
In what follows, we establish an effective non-linear field theory for this interacting interfacial dynamics, and probe it experimentally by applying statistical measures to the ensemble of experimental repetitions.

From the hydrodynamical equations governing the bulk motion of each fluid layer, one can derive the linear dynamics~\cite{Kumar94} of the interfacial height $\xi$ in terms of its spatial eigenmodes $\xi_{mk}(t)$. The latter are coefficients of an expansion of $\xi(t,r,\theta)$ in orthogonal eigenfunctions $f_{mk}(r,\theta)$ of the 2D Laplacian in polar coordinates~\cite{Ziener2015OrthogonalityFunctions}, with eigenvalues $-k^2$. For a fixed azimuthal number $m$, the radial confinement ($r_1\le r\le r_2$) in Fig.~\ref{fig:setup_evol} (e) discretises the infinite spectrum of positive wavenumbers $k$, resulting in a reduced density of states available at the interface; see Supplementary Information for a discussion of the boundary conditions. In order to obtain the non-linear dynamics of one of these modes $\xi_{mk}$, we invoke the variational formulation described by Miles~\cite{Miles1976NonlinearBasins,Miles1984NonlinearResonance}, and derive the relevant interaction terms that contribute to this single-mode evolution. By fixing the azimuthal number $m$, the approximate driven nonlinear equation of motion of a single mode $k$ reads
\begin{equation}
    \ddot{\xi}_{mk}+\left(2\gamma_k + \tilde{\gamma}_k[\xi]\right) \dot{\xi}_{mk}+\left(\omega_k^2(t)+\tilde{\delta}_k[\xi]\right)\xi_{mk}=\tilde{\eta}[\xi],\label{eq:stochasticDynamics}
\end{equation}
where
\begin{equation}
    \omega_k^2(t) = \frac{(\rho_1-\rho_2)(g-F(t))+\sigma k^2}{\rho_1+\rho_2}k\tanh(kh_0), \label{eq:dispersion}
\end{equation}
and the amplitude-dependent nonlinear terms $\tilde{\gamma}_k$, $\tilde{\delta}_k$ and $\tilde{\eta}$ are derived in Supplementary Information. In Eq.~\eqref{eq:stochasticDynamics} the linear damping $\gamma_k$, which encompasses any viscous contributions, is a phenomenological addition to the predicted dynamics, as discussed in Kumar \& Tuckerman~\cite{Kumar94}.

At first, we benchmark our proposed model of Eq.~\eqref{eq:stochasticDynamics} against the experimental results by simulating the single-mode evolution with the inclusion of self-interaction terms only. Due to its dominant unstable growth, the azimuthal mode $m=4$ reaches amplitudes at least one order of magnitude larger than the remaining modes and, thus, receives negligible contributions from them (see Supplementary Information). Thus, the approximate nonlinear terms are given by

\begin{subequations}
\begin{align}
    \tilde{\gamma}[\xi] &\approx -\gamma_k A_{k}\xi_{mk}^2k\tanh(kh_0), \\
    \tilde{\delta}[\xi]&\approx \frac{1}{2} A_{k}(\dot{\xi}_{mk}^2-\omega_k^2(t)\xi_{mk}^2) k\tanh(kh_0),
\end{align}
\label{eq:effnonlin}
\end{subequations}
with the numerical coefficient $A_{k}$ defined in Eq.~\eqref{eq:A4} of Supplementary Information for $p\equiv p'\equiv k'\equiv k$. By disregarding interactions with different modes, the source term $\tilde{\eta}[\xi]$ in Eq.~\eqref{eq:stochasticDynamics} only exhibits a stochastic noise term $\eta_k(t)$. This quantity accounts for the dynamics at microscopic scales, where the interface jitters due to the influence of random molecular (Brownian) motion~\cite{gardiner2004handbook} and environmental noise sources. Consequently, as each experimental repetition starts, those interfacial fluctuations set a stochastic initial state that later on evolves in a practically deterministic way. Linear evolution preserves the initial distribution for the parametrically amplified interfacial fluctuations. This picture is consistent with the experimental ensemble of amplitudes shown in Fig.~\ref{fig:setup_evol} (f) by the light blue lines, where we can see all repetitions growing uniformly until amplitude plateaus appear.

\begin{figure}
    \centering
    \makebox[0pt]{ \includegraphics[width = 1.\linewidth]{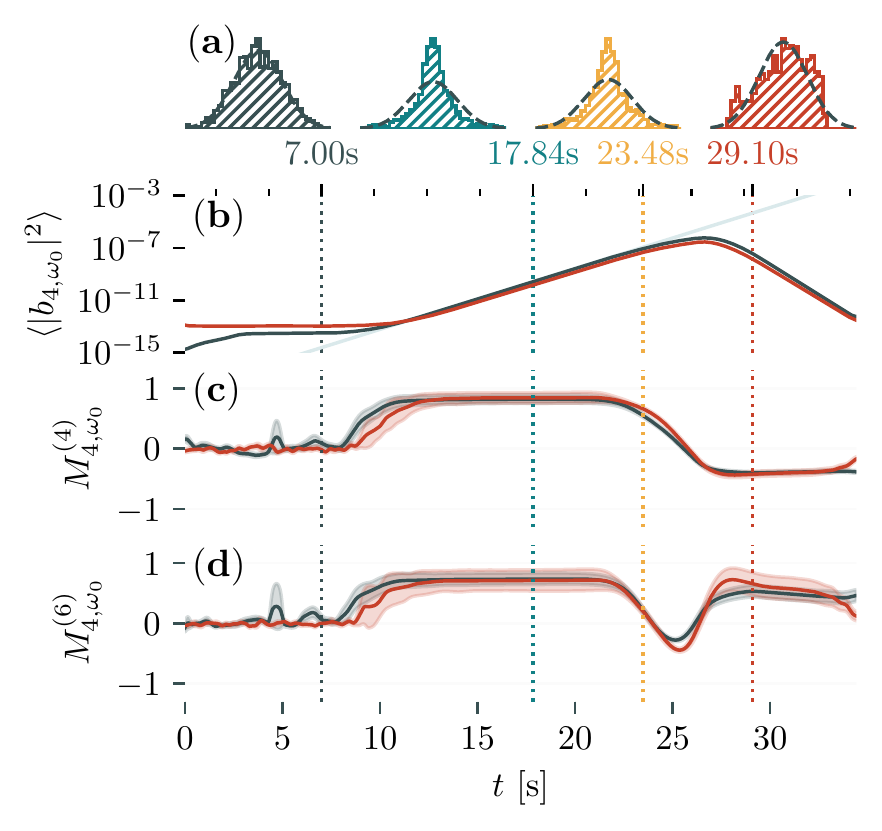}}
    \caption {\textbf{Ensemble properties of primary instability.}
    (\textbf{a}): Ensemble distributions of (the real part of) the amplitudes $b_{m,\omega_0}(t)$ of the $m=4$ mode at four different times (vertical dotted lines): the noise-floor (dark-green), the log-linear unstable growth (turqoise), the non-linear period (orange) and the decay (red). (\textbf{b}): The averaged squared amplitudes $\langle |b_{4,\omega_0}|^2\rangle$ for experiment (dark green) and simulation (red). (\textbf{c}) and (\textbf{d}) depict the statistical measures $M^{(2n)}_{m,\omega}$ for $n=2,3$ computed on both experimental data (dark green) and simulations (red), with bootstrapped bands of one standard deviation. Overall, we observe agreement between the self-interacting model and the experimental results.
    }
    \label{fig:cumulants}
\end{figure}

In Fig.~\ref{fig:cumulants} (a), we display the ensemble distributions of the instantaneous amplitudes $b_{4,\omega_0}$ at the four stages of experimental repetitions illustrated in Fig.~\ref{fig:setup_evol} (f). As expected from white detection noise in the first region (at $7$ seconds), we observe normally distributed amplitudes, which then evolve to more intricate, non-Gaussian distributions at later times.
The average squared amplitude, $\langle|b_{4,\omega_0}|^2\rangle$, is shown in Fig.~\ref{fig:cumulants} (b), for both the experimental (dark-green solid line) and simulated (red solid line) ensembles. This quantity confirms that our effective model accurately predicts the average amplitude of the dominant nonlinear unstable mode throughout the entire evolution.
To characterise the deviation from a featureless Gaussian and identify the onset of non-linearities, we employ a set of convenient measures of non-Gaussianity~\cite{Schweigler2017Experimental}, corresponding to equal-time correlators in field theories. These are defined in terms of higher-order statistical moments $\langle\cdot\rangle$ and cumulants $\langle\cdot\rangle_\mathrm{c}$ of the amplitudes $b_{m,\omega}$, see Supplementary Information, as follows
\begin{equation}
    M^{(2n)}_{m,\omega}(t,r_0) \equiv \frac{
    \langle \left( b_{m,\omega}^* b_{m,\omega}\right)^n \rangle_\mathrm{c}
    }{
    \langle \left( b_{m,\omega}^* b_{m,\omega}\right)^n \rangle}~.
    \label{eq:Gamma_correlation_def_appendix}
\end{equation}
For a normally distributed classical ensemble, the numerator vanishes at all orders of $2n$ greater than $2$, and, hence, the quantity vanishes entirely~\cite{marcinkiewicz1939summability}. The denominator of the above equation is commonly referred to as the full correlation function, while the numerator is its connected part. The latter vanishes for non-interacting fields for $n>1$, and it is the fundamental quantity for computing particle scattering and decay processes~\cite{peskin2018introduction,zinn2021quantum}. In Fig.~\ref{fig:cumulants} (c) and (d), solid dark-green lines display the non-Gaussianity measures $M^{(2n)}_{m,\omega}$ for the mode $m=4$, at the primary resonance frequency $\omega_0$, at even orders $n=2$ and $n=3$. As expected, the experimental ensemble results in nearly vanishing non-Gaussianity measures in the white noise-dominated region (before $8$ seconds). We recover this property in our model by introducing a Gaussian noise floor in our simulated ensemble, whose results are shown by the red curves in Fig.~\ref{fig:cumulants}.

As the signal of the repetitions leaves the background noise, all curves ramp up and reach a stable non-zero value, indicating an initial non-Gaussian distribution of interfacial waves. The constant value between $10$ and $21$ seconds shows the expected linear evolution of the ensemble at sufficiently small amplitudes. At around $22$ seconds, both measures experience abrupt changes, demonstrating a deviation from the linear distribution-preserving evolution of the ensemble. We see that, before the time indicated by the yellow dashed line in Fig.~\ref{fig:cumulants}, the simulated and experimental curves match for $M_{m,\omega}^{(4)}$ and $M_{m,\omega}^{(6)}$ within the $1\sigma$-confidence intervals, shaded regions in Fig.~\ref{fig:cumulants}. For both non-Gaussianity measures, we observe a quantitative discrepancy between simulation and experiment in the non-linear region.
Nevertheless, our model shows that the non-linearity is dominated by the self-interaction of the dominant parametrically amplified mode. The observed nonlinear dynamics signals the transition to the nonlinear Faraday resonance which may lead to stationary pattern formation on the interface in certain configurations in the long time limit~\cite{Edwards1994PatternsExperiment,Zhang1996SquareWaves,Zhang1997PatternWaves}. Here, we are interested in the onset of these nonlinearities prior to the saturation of the amplitudes. In this regime, a limited number of scattering channels are determined by the dominant parametrically amplified mode, leading to secondary instabilities with higher wavenumber, in correspondence with preheating dynamics in the early universe.

\begin{figure}[t]
    \centering
    \makebox[0pt]{ \includegraphics[width = 1.\linewidth]{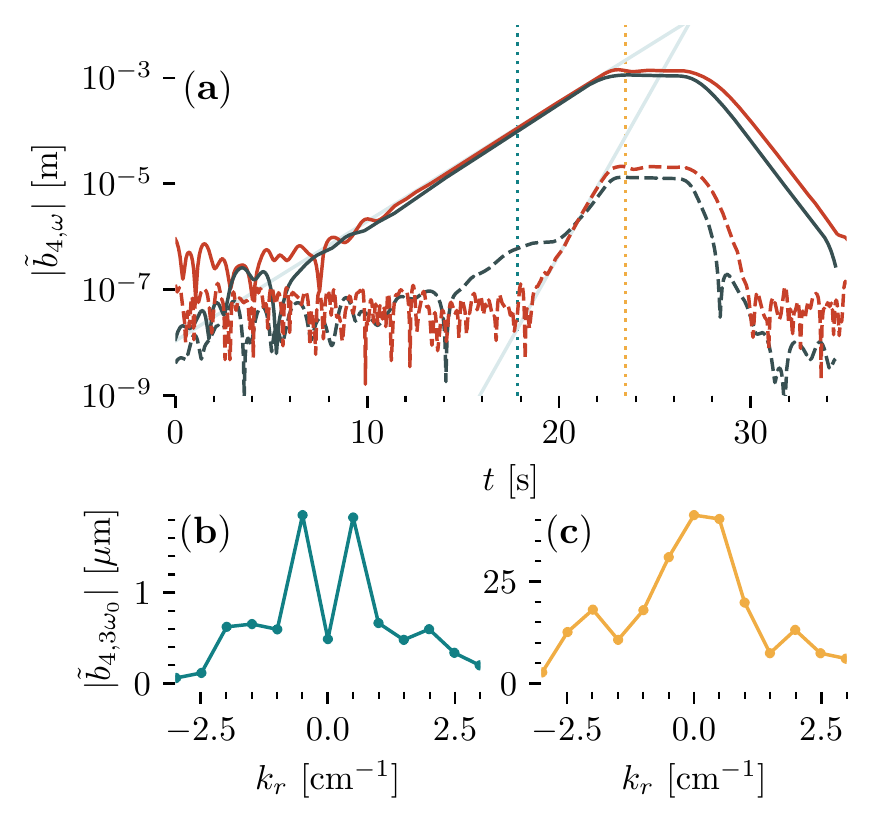}}
    \caption {
        \textbf{Thermalisation by secondary instabilities.}
        (\textbf{a}): Radially Fourier transformed instantaneous amplitudes, $\tilde{b}_{4,\omega}(t,k_r)$,  at various $\omega$ and $k_r$, of experimentally reconstructed (dark-green lines) and simulated (red lines) data.
        The primary instabilities at $(\omega_0,k_0)$ are displayed as solid lines, and the secondary instabilities at $(3\omega_0,k_1)$ are depicted by dashed lines.
        Solid light blue lines indicate the log-linear fit of the growth region of the simulated $|\tilde{b}_{4,\omega_0}(t,k_0)|$ and $|\tilde{b}_{4,3\omega_0}(t,k_1)|$, with growth rates $\lambda_0=0.52~\mathrm{s}^{-1}$ and $\lambda_1=1.48~\mathrm{s}^{-1}\approx 2.85\lambda_0$, respectively, verifying the $3:1$ ratio predicted by preheating mechanism (see main text for details).
        (\textbf{b}) and (\textbf{c}) show the radial spectra of the experimental amplitudes $\tilde{b}_{4,3\omega_0}(t,k_r)$ at two different times, matching the colors of the dotted vertical lines in (a). This verifies the thermalisation mechanism of preheating, i.e., the broadening of primary instabilities at low $k_0$ and the nonlinear amplification of the secondaries at higher $k_1$.
    }
    \label{fig:secondaries}

\end{figure}

So far we have only considered a self-interacting truncated model. This reduces the complete effective Lagrangian (Eq.~\eqref{eq:nonlinLag} in Supplementary Information) to include only non-linear terms proportional to $(\xi_{m,k}\dot{\xi}_{m,k})^2$, with a resulting equation of motion that does not allow the amplification of any other modes. Going beyond self-interactions, the dominant primary Faraday instability $\xi_{4,k_0}$ can source the dynamics of a secondary mode $\xi_{l,k_1}$, which is captured by including interaction terms proportional to $\xi_{l,k_1}\xi_{4,k_0}\dot{\xi}_{4,k_0}^2$ and $\xi_{4,k_0}^2\dot{\xi}_{4,k_0}\dot{\xi}_{l,k_1}$ in our model. In these terms, we note that modes with azimuthal number $l=4$ exhibit relatively low damping and large non-linear coefficients. In this case, the only remaining degree of freedom is the wave-number $k_1$. Therefore, accompanying the primary instability (solid dark line in Fig.~\ref{fig:secondaries}(a)) at $m=4$ with $k_0/(2\pi)\approx 0.35~\mathrm{cm}^{-1}$, we observe the growth of the secondary mode $l=4$ with $k_1/(2\pi)\approx 1.77~\mathrm{cm}^{-1}$ at the $3\omega_0$-resonance band (dashed dark line). These quantities are displayed in Fig.~\ref{fig:secondaries} (a) as the radial Fourier transform $\tilde{b}_{4,\omega}(t,k_r)$ of the instantaneous amplitudes $b_{4,\omega}(t,r)$ with radial wave-number $k_r$.

Berges \& Serreau~\cite{Berges2003ParametricTheory} employ approximate analytical and numerical techniques to identify a signature of preheating, arising from a $\phi^4$-type interaction, as the scattering of the primary to the secondary instabilities, where the latter appears with an integer multiple of the frequency and slope of the former. As outlined above, our system provides an analogue simulator for mode-mode scattering exhibiting quartic interactions, and as such grants a platform to investigate preheating experimentally. Our theoretical model (red lines in Fig.~\ref{fig:secondaries} (a)) accurately captures the non-linear features of the observed mode-mode interaction between primaries (solid lines) and secondaries (dashed lines), as depicted in Fig.~\ref{fig:secondaries} (a). We extract the slopes (plotted as light blue lines) for both primary and secondary instabilities, first for a single simulated run, obtaining the ratio $2.85$, and then for the entire simulated ensemble (see Extended Data Fig.~\ref{fig:secondarySlopes}), for which we obtain $3.06$.

The behaviour of the experimental secondary instability (dark dashed curve) in Fig.~\ref{fig:secondaries} (a) is due to an overlap between both low $k_0$ and high $k_1$ modes at frequency $3\omega_0$, resulting in a radial spectrum that can not be separated. Thus, when examining the slope of the experimental amplitude $\tilde{b}_{4,3\omega_0}(t,k_1)$, we observe a contribution from the primary growth to the pure secondary instability, which does follow the simulated model (red dashed curve). The relative contribution of the modes depends on their random initial state and hence varies between experimental repetitions. The outcome of this mode superposition is a damped secondary growth (see Extended Data Fig.~\ref{fig:slopeProblems}) preventing us from reliably comparing the extracted slopes to our model. Qualitatively, in Fig.~\ref{fig:secondaries} (b) and (c), we observe in our experiment the broadening of the primary resonance at $k_0$ and scattering into the secondary instability with higher $k_1$, as predicted in preheating models in Bose-Einstein condensates~\cite{Robertson2018NonlinearitiesCondensates,Butera2022}. Our findings support the preheating scenario and approximate techniques put forward by Berges \& Serreau~\cite{Berges2003ParametricTheory} and theoretical analogue preheating proposals in ultra-cold atoms~\cite{Zache2017InflationaryCondensates}. Additionally, the results presented here motivate the development of experiments and data analysis tools to reliably extract scattering amplitudes from analogue simulations.

By revisiting the century-old phenomenon of interfacial Faraday instabilities, we have outlined how carefully prepared, repeatable experiments can be used to simulate the key aspects of preheating. Our findings show that the mechanism of preheating prevails in our system despite the extra complications, such as dissipation and additional scattering channels. Our results are in support of universality and robustness of theoretical models tackling the thermalisation of the Early Universe and its distinct stages. As we have shown, analogue preheating simulators open a new avenue of investigation, with the potential to explore regimes beyond those we can calculate.

\putbib
\end{bibunit}

\noindent\textbf{Acknowledgments} SW is acknowledging Joerg Schmiedmayer's support and numerous eye-opening discussions on extracting effective field theories from hydrodynamical systems. SW and SE acknowledge discussions with Torsten Zache, whose insights on preheating simulations and encouraging feedback played an essential role in this project. The authors are grateful to Bill Unruh for his invaluable input in understanding the effect of temperature variations on the experimental setup. We thank Ed Copeland and David Kaiser for their detailed feedback on this work. The authors are grateful to Terry Wright, Pete Smith, Sionnach Devlin, Andrew Stuart and Tommy Napier for technical support. In particular, we are indebted to Terry Wright, whose expertise enabled the construction of the high-performance vertical acceleration platform. VB and ZF acknowledge several fruitful experimental discussions with Naresh Sampara and George Hunter-Brown.

SW acknowledges support provided by the Leverhulme Research Leadership Award (RL-2019- 020), the Royal Society University Research Fellowship (UF120112) and the Royal Society Enhancements Grant (RGF/EA/180286 and RGF/EA/181015), and partial support by the Science and Technology Facilities Council (Theory Consolidated Grant ST/P000703/1), the Science and Technology Facilities Council on Quantum Simulators for Fundamental Physics (ST/T006900/1) as part of the UKRI Quantum Technologies for Fundamental Physics programme. RJAH acknowledges support provided by the Leverhulme Trust (RPG-2018-363). SE acknowledges support through the EPSRC Project Grant (EP/P00637X/1) and an ESQ (Erwin Schr\"odinger Center for Quantum Science and Technology) fellowship funded through the European Union’s Horizon 2020 research and innovation program under Marie Skłodowska-Curie Grant Agreement No 801110. This project reflects only the author’s view, the EU Agency is not responsible for any use that may be made of the information it contains. ESQ has received funding from the Austrian Federal Ministry of Education, Science and Research (BMBWF).

\noindent\textbf{Author contributions}
VB, AG, ZF and SE performed the experiment and the
data analysis. VB and AG did the theoretical calculations. SE, AA, RJAH and SW provided scientific guidance in experimental and theoretical aspects of this work.
AA, RJAH and SW proposed the analogy and designed the experiment. All authors contributed to interpreting the data and writing the manuscript.

\noindent\textbf{Supplementary Information} is available for this paper.

\clearpage
\makeatletter
\widetext
\makeatletter
\apptocmd{\thebibliography}{\global\c@NAT@ctr 0\relax}{}{}
\makeatother
\begin{bibunit}
\import{}{arxiv_SM.tex}
\end{bibunit}

\end{document}

%% file: arxiv_SM.tex
\makeatletter
\renewcommand{\fnum@figure}{\textbf{Supplementary Fig. \thefigure}}
\renewcommand{\theequation}{S\@arabic\c@equation}
\makeatother
\setcounter{figure}{0} 
\setcounter{equation}{0}

\section*{Supplementary Information}

\subsection{Methods}
\label{sec:Methods}
\par
\textbf{Fluids preparation and properties}
The biphasic solution of potassium carbonate, ethanol and water is prepared under controlled conditions to prevent contamination, and in large quantity, to accurately respect the mass fractions in line $2$ of Table $1$ of Salabat \& Hashemi~\cite{Salabat2007}. After mixing, the solution stratifies into an upper layer, with a predominantly ethanol-water organic phase and a lower one, mostly consisting of an aqueous potassium carbonate phase. Their measured densities are $\rho_2=907(7)~\mathrm{kg}~\mathrm{m}^{-3}$ and $\rho_1=1276(10)~\mathrm{kg}~\mathrm{m}^{-3}$, respectively, and their uncertainties account for variations in the environmental temperature $T\approx 24-26^{\circ}\mathrm{C}$. The measured surface tension coefficient at the liquid-liquid interface using the pendant drop method~\cite{Hansen1991SurfaceAnalysis} is $\sigma=2.5(10)\times 10^{-3}~\mathrm{N}~\mathrm{m}^{-1}$. An engineered, sealed annular cylindrical cell with transparent windows at the top and bottom is filled using threaded Luer lock adapters and syringes. This enables precise control over the relative depth between fluid phases while preventing the formation of bubbles and other contaminants from entering.

\par
\textbf{Shaking platform and automation}
A bespoke platform oscillates the fluid cell vertically, guided by pneumatic bearings and suspended by metal springs. 
The spring-mass system is driven by a voice-coil actuator, controlled directly by the experimental computer.
The structure is built on levelling screws, which, together with accelerometer measurements, enable us to align the setup and monitor its performance during each repetition.
We set up $1500$ automated and synchronised independent repetitions of the driving loop taken across $49$ hours and $47$ minutes.
By appropriately setting their duration, we can observe Faraday instabilities while preventing the breaking of the interface and mixing of the fluid phases. Thus, for each repetition, the fluid cell oscillates for $160$ cycles at frequency $2\omega_0$, corresponding to $26.36 ~ \mathrm{s}$, and it rests for $93.6~\mathrm{s}$ before a new repetition starts. During the first $35~\mathrm{s}$, the interface is recorded so that its elevation $\xi$ is reconstructed from the experiment using a standard method of fluid profilometry, namely a two-dimensional variant of Fourier Transform Profilometry~\cite{Wildeman2018Real-timeBackdrop}, adapted and optimized to work with large optical occlusions while minimizing the effect of local errors.

An accelerometer measures the overall cartesian acceleration $(a_x,a_y,a_z)$ experienced by the platform.  We reduce the combined horizontal component $a_r=\sqrt{a_x^2+a_y^2}$ to prevent off-axis sloshing of the fluids, which could lead to unwanted effects and jeopardize parametric resonance. The cross-axis ratio $a_r/a_z$ remained well below $0.5\%$ throughout the entire experiment, averaging at $0.396(4)\%$ (Cf.~\cite{Harris2015GeneratingBearing,ISOTransducer}), and the total harmonic distortion of the first $20$ harmonics of $2\omega_0$ stayed under $1.5\%$.  Variations in the laboratory's temperature prompted larger changes in the driver's acceleration. Regardless, these varying amplitudes stay within $\pm 1.5\%$ of the average at $a_z = 3.45(3) \ \mathrm{m\ s^{-2}}\equiv F_0$ and are not large enough to jeopardise the experiment. 

\textbf{Numerical simulations:}~
The simulations exhibited in Figures~\ref{fig:cumulants} and~\ref{fig:secondaries}, are obtained through numerical simulation of equation \eqref{eq:stochasticDynamics} using the effective non-linear terms presented in equations \eqref{eq:effnonlin}. These simulations consist of synthetically generated, discrete values $\xi_{n}^{(j)}=\xi_{km}^{(j)}(n\Delta t-\Delta t)$, where $n=1,...,N_t$ are the timesteps, $j=1,...,1500$ the realisations and $(N_t-1)\Delta t=35$ seconds with $N_t=3000$. For each simulation $j$, the time-dependent dispersion frequency $\omega_k(n\Delta t)$ is calculated using Eq.~\eqref{eq:dispersion} with fluid parameters presented here, where $F(t)$ is taken to be the synchronized, measured acceleration of the platform for the $j$'th experimental run. The eigenfrequency $\omega_0/(2\pi)=3.035~\mathrm{Hz}$ is the same across all realizations $j$, and is computed using Eq.~\eqref{eq:dispersion}, the measured fluid quantities and the wavenumbers $k_0/(2\pi)\approx 0.35~\mathrm{cm}^{-1}$ and
$k_1/(2\pi)\approx 1.77~\mathrm{cm}^{-1}$, obtained from the analytic radial boundary conditions~\cite{Ziener2015OrthogonalityFunctions}(see Supplementary Information). 

Starting with a vanishing amplitude $\xi_{km}(t=0)=0$ at rest, i.e. $\dot{\xi}_{km}(t=0)$=0, the time evolution is obtained by performing a 4th order Runge-Kutta finite difference scheme for the deterministic part ($\tilde{\eta}=0$). The stochastic contribution $\Delta\tilde{\eta}$ during the time interval $\Delta t$ is taken to be real gaussian white noise, and is applied after each deterministic timestep. That is, we write $\Delta \tilde{\eta}(t)=N_j(t)\Delta t$, where $N_j$ is real gaussian noise with zero mean and standard deviation $\sigma_j$. 
Where, if the noise is entirely of thermal origin, we expect a fluctuation-disspation relation~\cite{gardiner2004handbook} of the form $\sigma_j^2 \propto \gamma_k^{(j)}$, where $\gamma_k^{(j)}$ is the estimated damping $\gamma_k$ from the $j$'th experiment. Choosing to remain agnostic about the origin of the noise, we instead take 
\begin{equation}
    \sigma_j = \sigma_0 \frac{\xi_{km}^{(j)}(t_0)-\overline{\xi_{km}}(t_0)}{\sigma_\xi(t_0)}
    \label{eq:app:stds}
\end{equation}
where $\xi_{km}^{(j)}(t_0)$ is the measured amplitude in the $j$'th experiment at some time $t_0$ during the linear regime, and $\overline{\xi_{km}}(t_0)$ and $\sigma_\xi(t_0)$ is the (ensemble) mean and standard deviation of $\xi_{km}^{(j)}(t_0)$ respectively. The consequence of the distribution of $\sigma_j$ is to raise/lower the constant value between $12$ and $20$ seconds in Fig.~\ref{fig:cumulants} (b). Therefore, the choice \eqref{eq:app:stds}, along with the numerical value $\sigma_0 = 570\mathrm{nm}$, amounts to matching the initial non-Gaussianity in the simulation to the what is observed during the linear evolution. 

To mimic measurement noise, we add (central) Gaussian noise $s^{(j)}(n\Delta t)$ with standard deviation $\sigma=1.4\mathrm{\mu m}$ -- this value is taken from the experimentally observed noise-floor -- to the simulation result $\xi_{n}^{(j)}$, i.e. $\xi_{n}^{(j)}\mapsto \xi_{n}^{(j)} + s^{(j)}(n\Delta t)$. To obtain the complex amplitudes $b_{m,\omega}$, the simulated data $\xi_{n}^{(j)}$ undergoes the same post-processing as the reconstructed, experimentally observed amplitudes $\xi_{km}^{(j)}(t)$.

\par 
\textbf{Post processing:}
Following the experimental procedure outlined here, we are left with 1500 realisations of the interfacial height $\xi^{(j)}_{a,b,c}\equiv \xi^{(j)}(t_a,r_b,\theta_c)$ on a linearly spaced discrete polar mesh $(t,r,\theta)\in\mathbb{R}^{3443}\times\mathbb{R}^{128}\times\mathbb{R}^{256}$ with $r\in [22.1\mathrm{mm},37.9\mathrm{mm}]$. Unless otherwise stated, we select a single radius $r_0 = r_{16} = 24\mathrm{mm}$, which is chosen to be well separated from both the meniscus and the radial zero-crossing of the resonant mode. The complex amplitudes $\xi_m(t,r_0)$ are found from applying a Fast-Fourier Transform (FFT) along the azimuthal ($\theta$) axis. 
To perform the decomposition in instantaneous amplitudes $b_{m,\omega}$, we apply a cosine filter, centred at the frequency $\omega$ and with width $\Delta\omega = \omega/2$, to the temporal FFT of $\xi_m$, followed by an inverse FFT. The result are complex amplitudes $b_{m,\omega}(t,r_0)$ (for $\omega > 0$) and $b_{-m,\omega}(t,r_0)=b_{m,-\omega}^*$ (for $\omega < 0$). 
Note that nowhere in the processing has a filter in the mode number $k$ been applied. Instead, we note that for the azimuthal mode in question, i.e. $m=\pm 4$, the radial profiles have a single zero-crossing. Note that if $\omega$ is the only frequency present in $\xi_m$, then $2b_{m,\omega}(t)$ acts as the Hilbert transform, or analytic extension, of the real field $\xi_m+\xi_{-m}^*$. Due to the restricted reconstruction area in the radial direction, we are limited to performing a Fourier instead of a Bessel decomposition. This leads to the overlap of radial modes as discussed in the main text. 

\par 
\textbf{Computing the non-linear coefficients:}
The coefficients for the effective non-linear terms presented in \eqref{eq:effnonlin} are determined by the quantity $T_k A_{kkkk}$ derived in the Supplementary Information. To compute $A_{kkkk}$ we follow the approach of Ziener \cite{Ziener2015OrthogonalityFunctions} to obtain the sets of valid $k$ for each azimuthal number $m$. The coefficients \eqref{eq:app:coeffsCandD} are found using numerical integration, with $f_{km}$'s determined by the estimated $k$'s. Using equation \eqref{eq:A4}, and truncating the sum over $q$ by $m=-30,...,30$ with the $15$ lowest values for $k$ for each $m$ included, we obtain the final result $T_k A_{kkkk}=616309~\mathrm{rad}^2\cdot\mathrm{m}^{-2}$.

\subsection{Non-linear interfacial dynamics}
\label{app:nonlin}

\noindent
\par
\textbf{Spatial eigenfunctions, Floquet predictions and damping}
The interface $z=\xi(t,r,\theta)$ is decomposed into eigenfunctions $f_{mk}(r,\theta)$, with eigenvalue $-k^2$, of the 2D laplacian in polar coordinates
\begin{equation}
    \xi(t,r,\theta) = \sum_{m\in \mathbb{Z}}\sum_{k} \xi_{mk}(t)f_{mk}(r,\theta),
\end{equation}
for $f_{mk}(r,\theta)\equiv R_m(kr)\cos(m\theta)$, where $m$ is the azimuthal number. Imposing Neumann boundary conditions at the vertical walls $r=r_1$ and $r=r_2$, we find that the radial functions $R_m$ must be of the form
\begin{equation}
R_m(kr)=Y_{m}^{\prime}\left(k r_{2}\right) J_{m}\left(k r\right)-J_{m}^{\prime}\left(k r_{2}\right) Y_{m}\left(k r\right),
\end{equation}
where the discrete set of permitted $k$-values is determined by the condition
\begin{equation}
Y_{m}^{\prime}\left(k r_{2}\right) J_{m}^{\prime}\left(k r_{1}\right)=J_{m}^{\prime}\left(k r_{2}\right) Y_{m}^{\prime}\left(k r_{1}\right).
\end{equation}
Here, $J_m$ and $Y_m$ are the Bessel functions of the first and second kind, respectively, and $^\prime$ denotes derivative with respect to the argument. These radial functions further enjoy the symmetry $R_{m}(kr) = R_{-m}(kr)$, and the mode functions $f_{km}$ satisfy the following orthogonality condition
\begin{equation}
    (f_{km},f_{k'm'})\equiv\int d^2x f_{km}(x)f_{k'm'}(x')
    = \int_{-\pi}^{\pi} d\theta \cos(m\theta)\cos(m'\theta)\int_{r_1}^{r_2} dr ~ r R_m(kr)R_{m^\prime}(k'r)
    =\pi \mathcal{N}^2_{mk}\delta_{m,m'}\delta_{k,k'},
\end{equation}
where $\mathcal{N}_{mk}$ is a standard normalisation constant defined in equation $(68)$ of Ziener et al.~\cite{Ziener2015OrthogonalityFunctions}.

\begin{figure}[t!]
    \centering
    \makebox[0pt]{ \includegraphics[width = .6\linewidth]{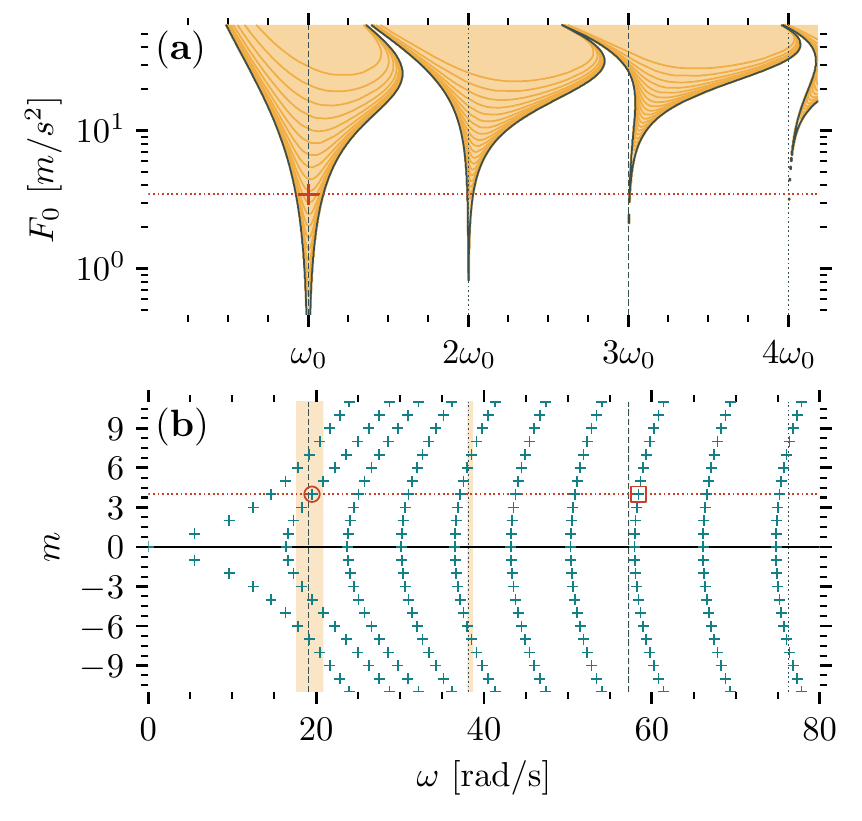} }
    \caption {
    (\textbf{a}): Instability regions corresponding to a given shaker amplitude $F_0$ are shaded in orange, with lines (thin orange) indicating how the region changes for logarithmically spaced values of damping $\gamma$. The vertical dashed lines indicate odd multiples of half the driving frequency, while the vertical dotted lines indicate the even multiples. The horizontal red dotted line highlights the value $F_0$ for the shaker amplitude in our setup.
    (\textbf{b}): The eigenfrequencies $\omega$ of the interface waves for different azimuthal numbers $m$ (turquoise crosses). For a given $m$, the frequencies can be counted, from the left, by the number $n$ of radial zero-crossings. Faint yellow bands are drawn to highlight the maximal expected instability region predicted from panel (a). Two $m=4$ modes (horizontal red dotted line) are indicated: the resonant mode at one zero-crossing (red circle), and the non-linearly populated mode at seven zero-crossings (red square).
    }
    \label{fig:instabilityRegions}
\end{figure}

\begin{figure} [ht]
\centering
\makebox[0pt]{ 
    \includegraphics[width = .45\linewidth]{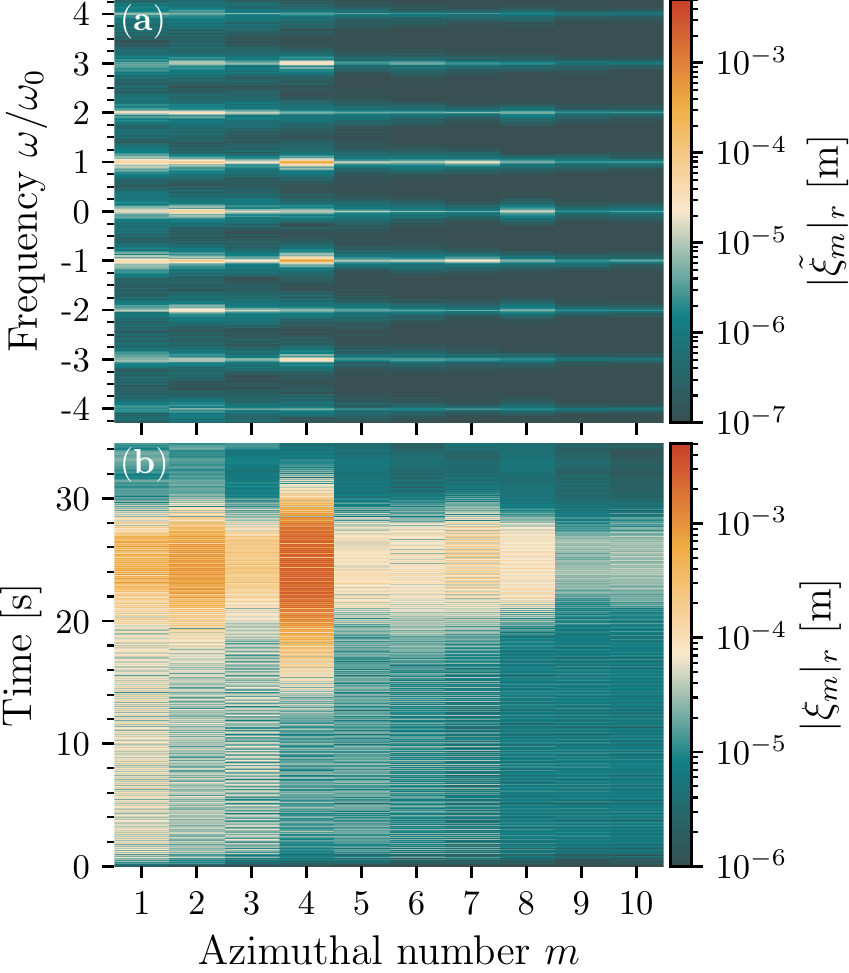}
}
\caption{
(a): Representative frequency spectrum of the first ten azimuthal modes $\xi_m$ averaged over the radial direction for one of the experimental repetitions, i.e., $|\tilde{\xi}_m|_r(\omega)\equiv \sqrt{\langle|\mathcal{F}_t \mathcal{F}_\theta\xi(t,r,\theta)|^2\rangle_r}$ with $\langle\cdot\rangle_r$ denoting the average over $r$. 
(b): Time evolution of the azimuthal modes $\xi_m$ in (a) averaged over the radial direction for one of the experimental repetitions, i.e., $|\xi_m|_r(\omega)\equiv \sqrt{\langle|\mathcal{F}_\theta\xi(t,r,\theta)|^2\rangle_r}$.
}
\label{fig:freq_aznum}
\end{figure}

Despite the menisci present in our system, the above approximation of exact Neumann boundary conditions at vertical walls yields an estimate for the radial wavenumbers $k$, following the numerical procedure of Ziener~\cite{Ziener2015OrthogonalityFunctions}. We then use these values to obtain predictions for linear Faraday interfacial instabilities, in line with Kumar \& Tuckerman~\cite{Kumar94}. The authors approximate the linear dynamics by a damped Mathieu equation with a phenomenological damping coefficient. Fig.~\ref{fig:instabilityRegions} displays the Floquet analysis instability predictions for the approximate linear dynamics, using the numerical routine proposed by Kovacic et al.~\cite{Kovacic2018MathieusFeatures} with several values of damping coefficient $\gamma$. In Fig.~\ref{fig:instabilityRegions}(b), we see the primary instability band around $\omega_0$ (shaded orange region), from which we conclude that modes with azimuthal numbers from $3$ to $8$ should undergo parametric amplification. It is worth noting that the menisci generate harmonic waves (at $2\omega_0)$~\cite{Douady1990ExperimentalInstability} and shift the instability thresholds of the modes~\cite{NguyemThuLam2011EffectThreshold}, causing discrepancies between the linear model assuming Neumann boundary conditions and the experimental results (see Fig.~\ref{fig:freq_aznum}).

It is often the case in studies of Faraday instabilities, that the experimentally measured damping is poorly described by phenomenological models only taking into account the dissipation of mechanical energy and disregarding the boundaries~\cite{Landau2013Fluid6,Kumar94}, see e.g.~\cite{Case57,Mei73,Miles90,Henderson1994Surface-waveLine,Henderson1990Single-modeCylinders,NguyemThuLam2011EffectThreshold}. Our system is no different and, due to its size, boundary effects must not be neglected. In order to obtain a consistent phenomenological model for the linear damping coefficients neglecting capillary effects from the menisci, we extend the results of Case \& Parkinson~\cite{Case57}, as follows: 
\begin{multline}
    \gamma_k = 2\frac{\rho_1\nu_1+\rho_2\nu_2}{\rho_1+\rho_2}k^2
    +\frac{\rho_1\sqrt{\nu_1}+\rho_2\sqrt{\nu_2}}{\rho_1+\rho_2}\sqrt{\omega_{km}}\{\mathcal{C}_{mk}(r_1)+\mathcal{C}_{mk}(r_2)\}\\
    +\frac{\rho_1\sqrt{\nu_1}+\rho_2\sqrt{\nu_2}}{\rho_1+\rho_2}\frac{k\sqrt{\omega_{km}}}{\sqrt{2}}\left\{\frac{1}{\sinh(2kh_0)}+\frac{1}{2\tanh(kh_0)}\right\},
\end{multline}
where
\begin{equation}
   \mathcal{C}_{mk}(r_j) =  \frac{r_1\mathcal{N}_{mk}^2 R_m^2(kr_j)}{4\sqrt{2}}\left(1+\frac{m^2}{k^2 r_j^2}\right)\left\{1-\frac{2k h_0}{\sinh(2kh_0)}\right\}.
\end{equation}

\par
\textbf{Non-linear modelling}
Before embarking on the derivation of non-linearities, we note that an effective model for single-mode non-linearities can be constructed using intuitive dimensional analysis arguments. 
First, observe that when the amplitudes grow large compared to the depth of the fluids, the activated non-linearities must introduce an energy loss to the system to prevent further growth. Assuming the effect can be thought of as an effective non-linearity in the dynamics of a single-mode, the non-linear terms must introduce a term that scales with $\xi_{km}^2$ to the damping -- a linear dependence would decrease damping in the wave-crests. The two natural candidates for such an effect, respecting the phase-preservative nature of the dynamics, are cubic non-linear terms of the form $\xi_{km}^2 \dot{\xi}_{km}$ and $\dot{\xi}_{km}^2 \dot{\xi}_{km}$. For consistency, the two additional terms $\dot{\xi}_{km}^2 \xi_{km}$ and $\xi_{km}^2 \xi_{km}$ are considered. Exploiting the natural scales $\omega_0$ and $k$, we may postulate coefficients using dimensional analysis. Then, finally, we are left with four non-linear terms, two are responsible for non-linear damping $k^2 \omega_0^{-1} \dot{\xi}_{km}^2 \dot{\xi}_{km}$ and $k^2 \omega_0 \xi_{km}^2 \dot{\xi}_{km}$, and two represent a de-tuning of the natural frequency $k^2 \omega_0^2 \xi_{km}^2 \xi_{km}$ and $k^2 \dot{\xi}_{km}^2 \xi_{km}$. Although numerical simulation using these terms capture the essential features of the non-linear stage, a variational approach for estimating the non-linearities of the real-valued amplitude $\xi_{km}$ has already been proposed by Miles~\cite{Miles1984NonlinearResonance}. We note, however, that the resulting non-linear terms are in agreement with those expected from the argument above, with small changes in the coefficients, and with the absence of the $\dot{\xi}_{km}^2$ damping term.

In what follows, we denote the unique pair of mode-numbers $(k,m)$ by lower-case Latin letters, i.e., $\xi_{km}\equiv\xi_a$ and $(k,m)\rightarrow(k_a,m_a)$, for simplicity. We can extend Miles' variational approach~\cite{Miles1984NonlinearResonance} to obtain a non-linear Lagrangian, perturbatively expanded in powers of the two-fluid interfacial height modes $\xi_{a}$, as follows
\begin{equation} 
    L = \sum_a \frac{\rho_1+\rho_2}{2T_a}\left[\dot{\xi}^2_{a}-(\omega_{a}^2(t))\xi^2_{a}\right] 
    +\frac{\rho_1-\rho_2}{2}\sum_{abc} A_{cab}\xi_{c}\dot{\xi}_{a}\dot{\xi}_{b} 
    +\frac{\rho_1+\rho_2}{4}\sum_{abcd} A_{cdab}\xi_{c}\xi_{d}\dot{\xi}_{a}\dot{\xi}_{b},\label{eq:nonlinLag}
\end{equation}
with $T_a\equiv k_a\tanh(k_a h_0)$ and the inclusion of damping provided by Rayleigh's dissipation function~\cite{Miles1976NonlinearBasins}
\begin{equation}
    Q_0 = \sum_a \frac{\rho_1+\rho_2}{T_a}\gamma_a\dot{\xi}^2_{a}.
\end{equation}
The coefficients $A_{cab}$ and $A_{cdab}$ are defined in terms of the mode functions $f_{a}$ similarly to equations $(2.6-7)$ in Miles 1984~\cite{Miles1984NonlinearResonance}, i.e., 
\begin{subequations}
    \begin{align}
        A_{cab}    \equiv& \frac{C_{cab}}{2T_a T_{b}}\left(2T_a T_{b}+k_c^2-k_a^2-k_b^2\right),\\
        A_{cdab}  \equiv& -\frac{D_{cdab}}{T_a T_{b}}\left(T_a + T_{b}\right)+2\sum_e \frac{C_{cae}C_{dbe}}{T_e T_a T_{b}}\left(k_e^2+k_a^2-k_c^2\right)\left(k_e^2+k_b^2-k_d^2\right),\label{eq:A4}
    \end{align}
\end{subequations}
with 
\begin{subequations}
    \label{eq:app:coeffsCandD}
    \begin{align}
                C_{cab}    &= \frac{1}{\pi\mathcal{N}_{a}\mathcal{N}_{b}}\int d^2x f_{a} f_{b} f_{c},\\
                D_{cdab}  &= \frac{1}{\pi\mathcal{N}_{a}\mathcal{N}_{b}}\int d^2x f_{c} f_{d} \nabla f_{a}\cdot \nabla f_{b}. 
    \end{align}
\end{subequations}
Resorting to Lagrange's equations applied to the above system, one can obtain the nonlinear equations of motion for the modes $\xi_{a}(t)$ with all interaction terms up to the relevant order. 

We take a closer look at two particular cases, namely: a dominant unstable mode experiencing self-interactions, and a subdominant mode subject to interactions with the first only. We further assume all other nonlinear terms are negligible. 
Both cases can be modelled using the following strategy: Using the Lagrangian \eqref{eq:nonlinLag}, we construct the equations of motion for an arbitrary mode $\xi_a$. We then truncate the sum based on the observation that if $\xi_b$ is the dominant mode, then $\xi_b \gg \xi_c$ for any $c\neq b$. 
To leading order, the equation of motion for $\xi_a$ can be written
\begin{equation} 
     \ddot{\xi}_a + 2\gamma_a \dot{\xi}_a + \omega_a^2(t)\xi_a 
     + \frac{\tilde{\rho}}{2}T_a(2A_{bba}-A_{abb}) \dot{\xi}_b^2 
    + \tilde{\rho}T_a A_{bba}\xi_b \ddot{\xi}_b
    + \frac{1}{4}T_a \mathcal{A}_{ba} \xi_b \dot{\xi}_b^2
    +\frac{1}{4}T_a(A_{bbab}+A_{bbba}) \xi_b^2 \ddot{\xi}_b
    \simeq 0,
    \label{eq:modemodeSourcing}
\end{equation}
where
\begin{equation}
    \mathcal{A}_{ba} \equiv 
    2A_{bbab}+2A_{bbba}-A_{abbb}-A_{babb},
\end{equation}
and $\tilde{\rho}\equiv \frac{\rho_1-\rho_2}{\rho_1+\rho_2}$ is the (dimensionless) Atwood number
Equation \eqref{eq:modemodeSourcing} depicts a parametrically driven oscillator $\xi_a$ that is non-linearly forced by the oscillation of the dominant mode $\xi_b$. Since this is valid for any mode $a$, we may consider the special case $a=b$ to obtain the non-linear self-interaction terms affecting the evolution of the dominant mode $\xi_b$. That is 
\begin{equation} 
    \left(
    1+\tilde{\rho}T_b A_{bbb}\xi_b +\frac{1}{2}T_b A_{bbbb} \xi_b^2
    \right)\ddot{\xi}_b 
    + \left(
    2\gamma_a 
    + \frac{\tilde{\rho}}{2}T_b A_{bbb}\dot{\xi}_b
    \right)\dot{\xi}_b 
    + \left(
    \omega_b^2(t) 
    + \frac{1}{2}T_b A_{bbbb}\dot{\xi}_b^2
    \right)\xi_b
    \simeq 0
    \label{eq:modemodeSelf}
\end{equation}
From its definition in Eq.~\eqref{eq:app:coeffsCandD}, it is straightforward to show that $C_{bbb}$ vanish for all modes $b$ with $m_b\neq 0$, and hence we have $A_{bbb}=0$. By simplifying the equation above and perturbatively inverting the multiplicative coefficient of $\ddot{\xi}_b$, we obtain
\begin{equation} 
    \ddot{\xi}_b
    +  \left(
    2\gamma_a -\gamma_aT_b A_{bbbb} \xi_b^2
    \right) \dot{\xi}_b 
    + \left(
    \omega_b^2-\Omega_b^2 
    + \frac{1}{2}T_b A_{bbbb}\left[\dot{\xi}_b^2-\omega_b^2(t) \xi_b^2\right]
    \right)\xi_b
    \simeq 0,
    \label{eq:modemodeSelf}
\end{equation}
which has the form of Eq.~\eqref{eq:stochasticDynamics} upon the following identification
\begin{align}
    \tilde{\gamma}[\xi]
    &\equiv  -\gamma_{k} A_{kkkk} \xi_{mk}^2 k\tanh(kh_0), \\
    \tilde{\delta}[\xi]
    &\equiv  \frac{1}{2}A_{kkkk}\left[\dot{\xi}_{mk}^2-\omega_{k}^2(t) \xi_b^2 \right]k \tanh(kh_0),\\
    \tilde{\eta}[\xi]
    &\equiv 0.
\end{align}

We further note that, if the subdominant mode $a$ shares the same azimuthal number as $b$, i.e., $m_a=m_b=m$, but with different $k_a$, then Eq.~\eqref{eq:modemodeSourcing} also reduces to the form of Eq.~\eqref{eq:stochasticDynamics} by identifying
\begin{align}
    \tilde{\gamma}[\xi]
    &\equiv 0, \\
    \tilde{\delta}[\xi]
    &\equiv 0, \\
    \tilde{\eta}[\xi] 
    &\equiv -\frac{1}{4}T_a \mathcal{A}_{ba} \xi_b \dot{\xi}_b^2
    -\frac{1}{4}T_a(A_{bbab}+A_{bbba}) \xi_b^2 \ddot{\xi}_b.
\end{align}

\subsection{Statistical methods}
\label{app:stats}

Within the Quantum Field Theory (QFT) framework, the most fundamental quantity is the generating functional, or characteristic function,~\cite{peskin2018introduction}
\begin{equation}
    Z[J] \equiv \left\langle \exp{\left(i\int_\Omega J(s) X(s) ds \right)} \right\rangle
    \label{eq:partitionFunction}
\end{equation}
where $X(s)$ is a random variable with auxiliary function $J(s)$ for each state $s\in \Omega$, e.g. $J(s)X(s)ds=s_m(t)\xi_{km}(t)dt$ with $\Omega=\mathbb{Z}\times \mathbb{R}$. Here, the importance of the characteristic function $Z$ lies mainly in its ability to generate all statistical moments, or full correlation functions, through functional derivatives $\delta /\delta J$ with respect to the currents of interest, i.e.
\begin{equation}
    \langle X(s_1) ... X(s_n) \rangle \equiv (-i)^n \frac{\delta^n Z[J]}{\delta J(s_1) ... \delta J(s_n)}\Bigg|_{J=0}.
    \label{eq:moments}
\end{equation}
Likewise, the logarithm $\ln Z$ of the characteristic function generates all cumulants, also called $n$-point functions or (connected) correlation functions, through
\begin{equation}
    \langle X(s_1) ... X(s_n) \rangle_\mathrm{c} \equiv (-i)^n \frac{\delta^n \ln Z[J]}{\delta J(s_1) ... \delta J(s_n)}\Bigg|_{J=0}.
    \label{eq:cumulant}
\end{equation}
Consequently, the $n$th order cumulants can be expanded (see e.g.~\cite{gardiner2004handbook}) in terms of the $n$ first moments using
\begin{equation}
    \left\langle \prod_{j=1}^n X(s_j) \right\rangle_\mathrm{c} = \sum_{\pi \in \Pi_{n}}(-1)^{|\pi|-1} (|\pi|-1)! \prod_{B\in \pi} \left\langle\prod_{i\in B} X(s_i) \right\rangle,
    \label{eq:cumulants_def}
\end{equation}
where $\pi \in \Pi_{n}$ is a partition of $n$ elements into $|\pi|$ blocks, $B\in\pi$ is a block in the partition, and $i\in B$ is an element in the block. 

Since the characteristic function $Z$ factorizes over independent variables, cumulants $\llangle X_1 ... X_n \rangle_\mathrm{c}$, being the coefficients of the series expansion of $\ln Z$, vanish if, and only if, all the variables $X_1,...,X_n$ are independent. This means that if modes evolve independently, then the only non-zero $n$-point functions are those of equal modes at equal times, i.e. $\llangle |b_{m,\omega}(t)|^{n} \rangle_\mathrm{c}$. Because moments of increasing order generally increase in size, e.g. $\langle X^{2n} \rangle \geq \langle X^n \rangle^2$, the $n$-point functions scale with the $n$-moment. To remove this trivial scaling, we introduce quantities
\begin{equation}
    M^{(2n)}_{m,\omega}(t,r) \equiv \frac{
    \langle \left( b_{m,\omega}^* b_{m,\omega}\right)^n \rangle_\mathrm{c}
    }{
    \langle \left( b_{m,\omega}^* b_{m,\omega}\right)^n \rangle}~.
    \label{eq:Gamma_correlation_def_appendix}
\end{equation}
Note that rotational symmetry of the container results in $\xi_{km}$ having uniformly distributed (complex) phases across realisations, so that $\xi_{km}(t)$ are statistically central variables, i.e. $\langle \xi_{km}(t) \rangle=0$. 

\putbib

\newpage

\begin{figure}[ht!]
    \centering
    \makebox[0pt]{ \includegraphics[width = .8\linewidth]{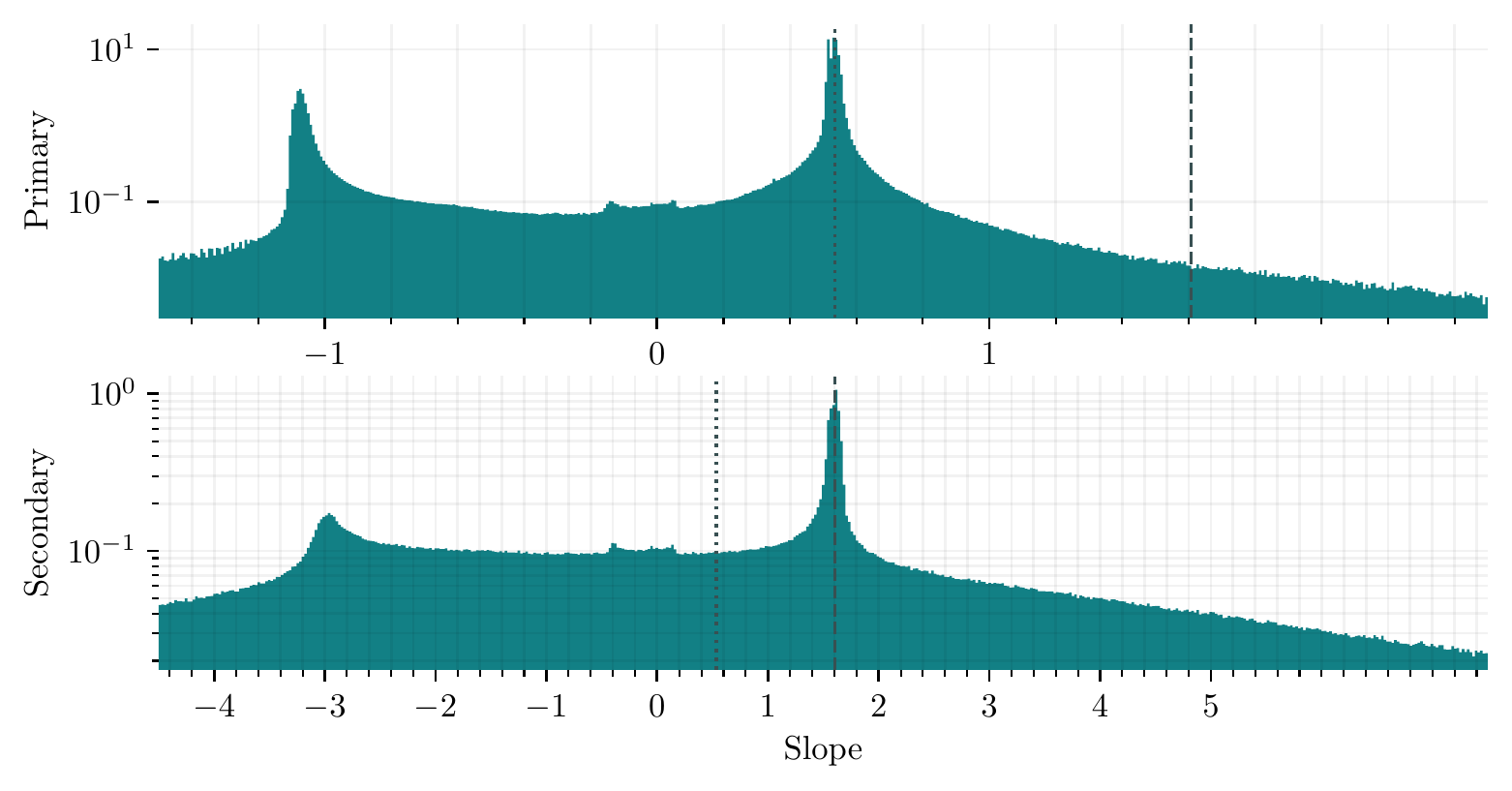} }
    \caption{\textbf{Histogram of slopes of simulated secondary and primary instabilities.}
    The instantaneous growth rates $\partial_t \log |b_{4,\omega}|$ in the simulation for the primary $k_0$ and the secondary $k_1$ across all realisations. The dotted vertical line signals the primary growth rate $\lambda$, the dashed vertical line signals the expected growth rate $3\times \lambda$ of the secondary. In both panels, the y-axis shows the normalised abundance of slopes across all realisations and times on logarithmic scale. 
    }
    \label{fig:secondarySlopes}
\end{figure}

\begin{figure}[ht!]
    \centering
    \makebox[0pt]{ \includegraphics[width = .8\linewidth]{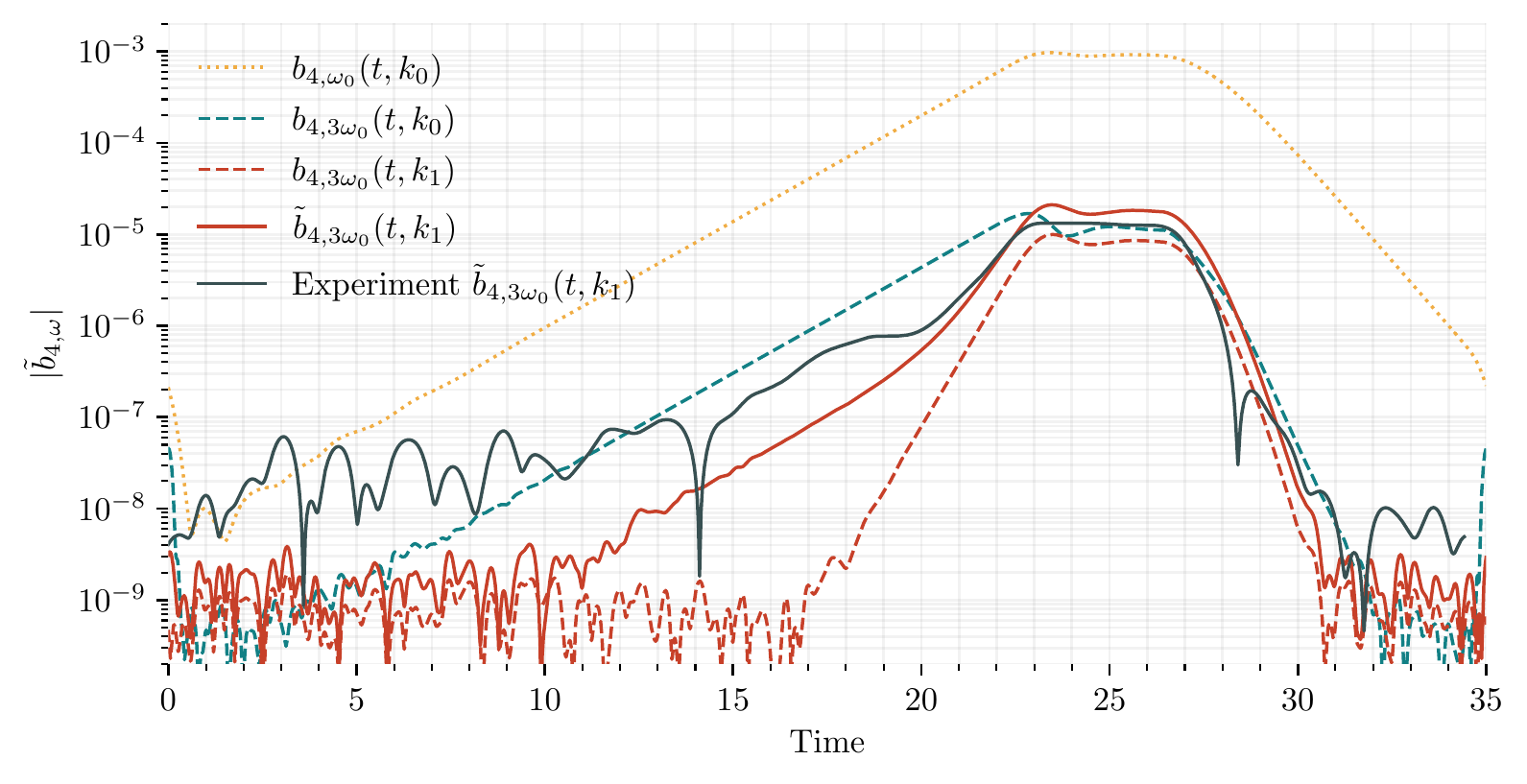} }
    \caption{\textbf{Simulated mode-decompositions.}
        The simulated logarithmic amplitude of the primary $b_{4,\omega_0}$ (orange dotted) is compared to the contribution of the primary to the $3\omega_0$-band (turquoise dashed) and the secondary (red dashed) $b_{4,3\omega_0}(t,k_1)$. The two simulated curves $b_{4,3\omega_0}(t,k_0)$ and $b_{4,3\omega_0}(t,k_1)$ are taken as coefficients of their respective radial modes $R_{m}(kr)$ to create a superimposed radial mode. This mode is filtered around $k_1$ using a radial fourier transform to obtain the curve $\tilde{b}_{4,3\omega_0}(t,k_1)$ (solid red). This is to be compared to the experimental curve (solid dark). The solid red curve demonstrates that the arbitrary composition of $b_{4,3\omega_0}(t,k_0)$ and $b_{4,3\omega_0}(t,k_1)$ results in a overall change in the secondary slope, evidenced by the solid red curve after $20$ seconds.
    }
    \label{fig:slopeProblems}
\end{figure}

\begin{figure}[ht!]
        \centering
        	\makebox[0pt]{ \includegraphics[width =.8\linewidth]{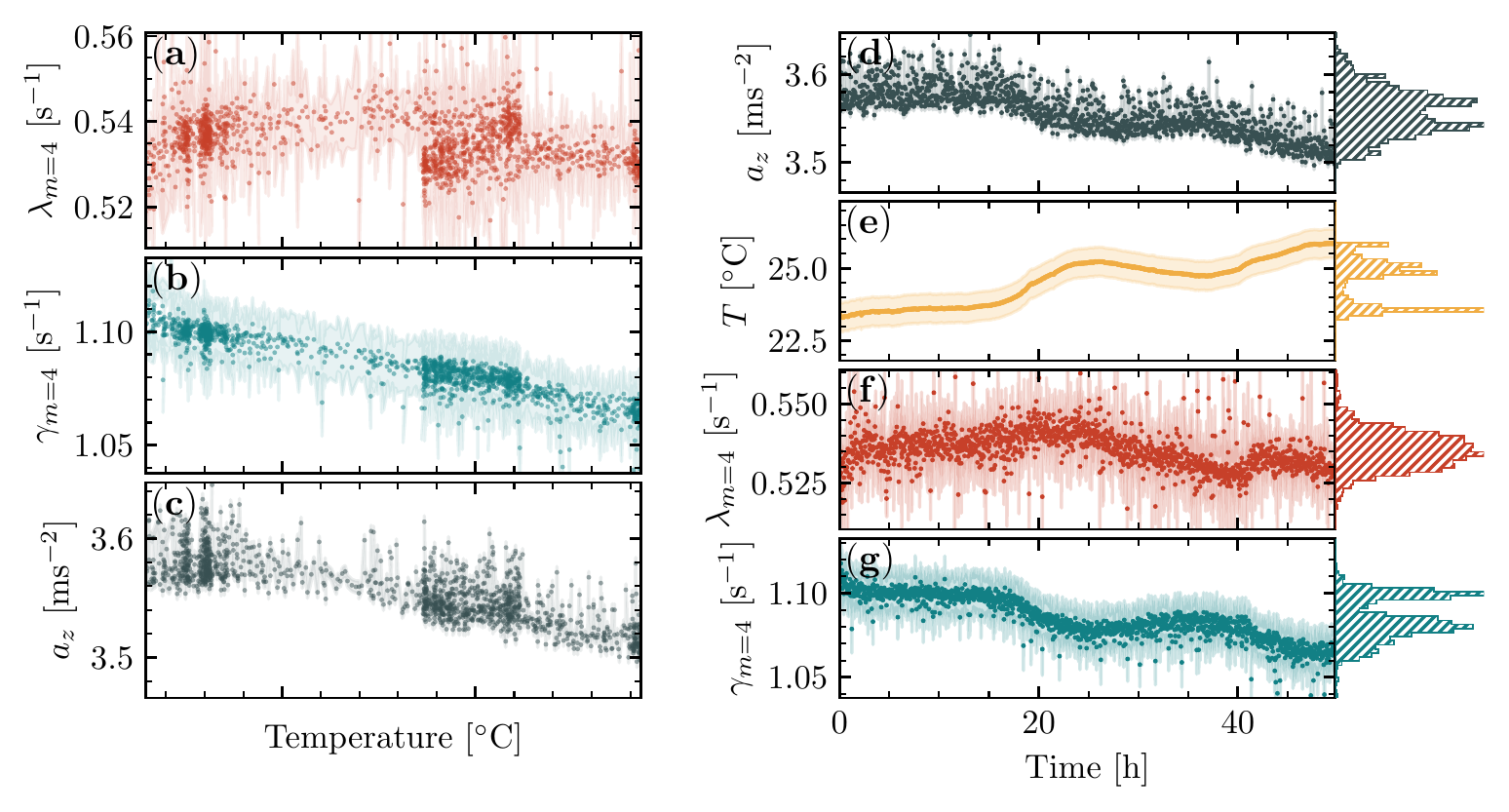} }
        	\caption {\textbf{Measured temperature and amplitude of vertical acceleration, and experimentally extracted growth and decay rates.}
        	\textbf{(a)}, \textbf{(b)} and \textbf{(c)} display the fitted growth and decay rates, and the vertical acceleration of the cell as functions of environmental temperature.
           \textbf{ (d)} shows the amplitude of the vertical driver acceleration $a_z$ measured throughout each run. \textbf{(e)} displays the average environmental temperature near the platform during each run.  Each run takes on average $119.5(19)$ seconds.
        	\textbf{(f)} and \textbf{(g)} display the fitted exponential amplification and decay rates for the azimuthal number $m=4$ around $\omega_0$,  $\lambda$ (blue dots) and $\gamma$ (red dots), respectively, by the elapsed time since the first experimental run. Their respective average values with uncertainty are $0.536(9) ~ \mathrm{s}^{-1}$ and $1.084(12) ~ \mathrm{s}^{-1}$. 
            The shaded regions indicate the uncertainty in the measured or fitted values.
        	}
        	\label{fig:scatter_all}
\end{figure}